\documentclass[nofootinbib,showpacs,preprintnumbers,amsmath,amssymb,floatfix]
{revtex4}
\usepackage{epsf}

\begin{document}


\title{Parametrization of the QCD coupling in the Evolution Equations}

\vspace*{0.3 cm}

\author{B.I.~Ermolaev}
\affiliation{H.~Niewodniczanski Nuclear Physics Institute PAN,
 31-342 Krakow, Poland \\ and Ioffe Physico-Technical Institute,194021
 St.Petersburg, Russia\footnote{Permanent address}}
\author{S.I.~Troyan}
\affiliation{St.Petersburg Institute of Nuclear Physics, 188300
Gatchina, Russia}

\begin{abstract}
We examine the parametrization of the QCD coupling in the Evolution Equations,
including DGLAP.
Our conclusion is that the well-known parametrization, where the argument of
the coupling is $k^2_{\perp}/\beta$ or just $k^2_{\perp}$, stands only if the lowest
integration limit in the transverse momentum
space (the starting point $\mu^2$
of the $Q^2$ -evolution) obeys the relation $\mu \gg \Lambda_{QCD} \exp {(\pi/2)}$,
otherwise the coupling should be replaced by the more complicated expression
presented in  Eq.~(\ref{aeff}).
\end{abstract}


\maketitle

\section{Introduction}
By various reasons, parametrization of $\alpha_s$ in QCD processes
has always been among the QCD topics on discussion.
In particular, the early summaries on the situation in the hard
processes were presented in
Refs.~\cite{av,ddt}  and one of the latest summaries is Ref.~\cite{ciaf}.
It is well-known (see e.g. Refs.~\cite{av,ddt}) that in the Feynman graphs
contributing to scattering amplitudes in the hard kinematics,
as well as to the Sudakov form-factors, the parametrization
of $\alpha_s$ is universal and simple:
\begin{equation}\label{aperp}
\alpha_s = \alpha_s (k^2_{\perp}),
\end{equation}
with $k_{\perp}$ being the transverse momentum of the soft gluon momenta.
This parametrization is also used in the DGLAP
equations in the integral form and leads to the parametrization
$\alpha_s = \alpha_s(Q^2)$ in the differential DGLAP equations.
Besides this example, there are other
cases where it is important to know the parametrization of $\alpha_s$.
For example, the parton distributions $F$ and some other objects
are often obtained through composing and solving the evolution equations of
the Bethe-Salpeter (BS) type, with one intermediate gluon factorized. Naturally,
such equations can be different and can be derived by different ways. In
particular, both DGLAP and BFKL equations belong to this group. We
do not consider obtaining those equations in the present paper.
Instead, our goal is to consider the treatment of $\alpha_s$ in BS equations in
general.
The parametrization of $\alpha_s$ in the BS equations was discussed in
Ref.~\cite{ddt}. In the much more detailed way the treatment of $\alpha_s$
in the BS equations was suggested
in Ref.~\cite{ds} where the parametrization
\begin{equation}\label{alfads}
\alpha_s = \alpha_s (-k^2_{\perp}/\beta),
\end{equation}
with $\beta$ being the longitudinal moment fraction
(see Eq.~(\ref{sud})), was derived.
In principle, the parametrization (\ref{alfads})  looks natural and
well-expected from the preceeding estimates, however the method of Ref.~\cite{ds}
includes serious contradictions which make the derivation of Eq.~(\ref{alfads}) unreliable.
In the present paper we  consider the parametrization of $\alpha_s$ in the Bethe-Salpeter
equations. We show what was the mistake done in Ref.~\cite{ds} and correct it.
As a result, we obtain a more complicated
expression for the effective coupling in the BS equations.  Then we demonstrate that
when the lowest limit
of integration over $k^2_{\perp}$ is large enough, the effective coupling
can be simplified down to $\alpha_s(k^2_{\perp}/\beta)$.
Earlier,
in Ref.~\cite{egtalpha} we argued that one of the easiest ways of treating $\alpha_s$
in the BS equation could be the use of the Mellin transform. Nevertheless by various
reasons, for example for Monte-Carlo simulations\cite{krak} and for performing the numerical
calculations in the way convenient for experimentalists\cite{was},
one may prefer not to involve this operation. To this end, in the present
paper we examine the treatment of $\alpha_s$ in the Bethe-Salpeter
equations without using the Mellin transform.
The paper is organized as follows: in Sect.~II we remind how the parametrization
of $\alpha_s$ in the hard kinematics, i.e. Eq.~(\ref{aperp}), can be obtained.
We suggest in Sect.~III the way to derive the parametrization of
$\alpha_s$ in the BS equations for the scattering amplitudes in the forward
kinematics. Due to the Optical theorem the
parton distributions are proportional to the imaginary parts of the forward
amplitudes. It allows us to fix in Sect.~IV the $\alpha_s$ -parametrization
in the BS equations for the parton
distributions. Finally, Sec.~V is for concluding remarks.

\section{Parametrization of $\alpha_s$ for QCD processes in the hard kinematics}

From the technical point of view, calculation of QCD processes in the
hard kinematics is much easier compared to the Regge kinematics, so to
begin with we remind known results for the parametrization of
$\alpha_s$ for the hard processes.
Let us consider the contribution $M_t$ of the Feynman graph depicted in Fig.~1.
It correspond to a certain factorization of the soft virtual gluon with momentum $k$
propagating in the $t$ -channel. The cases with $u$ and $s$ -channel gluons factorized can be
considered quite similarly.
In the present paper we do not discuss the factorization but
focus on the treatment of $\alpha_s$ only.
 The solid lines in Fig.~1 denote quarks, though
the generalization to the case of
gluons is obvious.
Through the paper we will assume that the lower particles, with momenta
$p_1,~p'_1$, have small virtualities $\sim \mu^2$ whereas virtualities of
the upper partons, with momenta $q,~q'$ are large:
$- q^2 \sim -q'^2\sim Q^2 \gg \mu^2$. This assumption allows us
to extend our analysis, with minor changes, to DIS and DVCS.
Applying the Feynman rules to Fig.~1 we obtain:
\begin{equation}\label{ms}
M_t = - \imath \int \frac{d^4 k}{(2 \pi)^4}
\frac{2 s}{[(q+k)^2 + \imath \epsilon][(p-k)^2 + \imath \epsilon]}
M(s, Q^2, (p-k)^2, (q+k)^2) 4 \pi \frac{\alpha_s (k^2)}{k^2 + \imath \epsilon}.
\end{equation}
Factor $2s$ in Eq.~(\ref{ms}) appears from simplifying the numerator. We have
denoted $s = (q+p)^2$ and assume that $\sqrt{s} \gg \mu$ and any of the
involved masses.
We dropped the color factors in Eq.~(\ref{ms}) as unessential for our analysis. $M$ corresponds to
the blob in Fig.~1.
In terms of the Sudakov variables
\begin{equation}\label{sud}
k = - \alpha (q +xp) + \beta p + k_{\perp}
\end{equation}
and therefore
\begin{equation}\label{sudinv}
k^2 = -s \alpha\beta - k_{\perp}^2,~~ 2pk = -s \alpha,~~ 2qk = s (\beta - x \alpha).
\end{equation}
In Eq.~(\ref{sudinv}) we have neglected the virtuality $p^2 = \mu^2$ of the
initial parton and denoted
$s = 2pq,~ x = Q^2/2pq,~Q^2 = - q^2$.
In terms of the Sudakov variables Eq.~(\ref{ms}) is
\begin{equation}\label{mssud}
M_t = \imath \frac{s^2}{4 \pi^2} \int d \alpha d \beta d k^2_{\perp}
\frac{M \big(s, Q^2,s\alpha, s \beta, k^2_{\perp}\big)}
{[s \beta - Q^2 - s \alpha\beta - k^2_{\perp} + \imath \epsilon]
[s \alpha- s \alpha\beta - k^2_{\perp} + \imath \epsilon]}
\frac{\alpha_s (- s \alpha\beta - k^2_{\perp})}{(s \alpha\beta + k^2_{\perp} - \imath \epsilon)}.
\end{equation}
Amplitude $M$ is unknown, so it is impossible to perform the integration
over any of the variables in Eq.~(\ref{mssud}). However, if we assume the
leading logarithmic (LL) accuracy, we can use the QCD -generalization\cite{fact}
of the bremsstrahlung Gribov theorem\cite{grib}. According to it, $M$ does
not depend on $\alpha$ and $\beta$. Integrating over $\alpha$ in Eq.~(\ref{mssud})
is conventionally performed with closing the integration contour down and taking
the residue at $s\alpha = (+ k^2_{\perp} - \imath \epsilon)/(1- \beta)$. It
converts Eq.~(\ref{mssud}) into
\begin{equation}\label{ahard}
M_t = - \frac{1}{2 \pi} \int_{\mu^2}^s \frac{d k^2_{\perp}}{k^2_{\perp}}
\int_{\beta_0}^1 d \beta\frac{(1- \beta)}{\beta} M(s,Q^2, k^2_{\perp})
\alpha_s \Big(- \frac{k^2_{\perp}}{(1-\beta)} \Big)
\end{equation}
where $\beta_0 = x + k^2_{\perp}/s$. Obviously, $\beta_0 \approx x = Q^2/2pq$
when $x \sim 1$ and the upper limit $s$ of the integration over
$k^2_{\perp}$ can be changed for $Q^2$.
The minus
sign of the $\alpha_s$ -argument in Eq.~(\ref{ahard}) indicates explicitly that
the argument is space-like and for the space-like argument $\alpha_s$ is given
by the well-known expression:
\begin{equation}\label{alphahard}
\alpha_s \Big(- \frac{k^2_{\perp}}{(1-\beta)} \Big) =
\frac{1}{b \ln \Big(k^2_{\perp}/\big((1-\beta)\Lambda^2\big)\Big)}.
\end{equation}
We have used here the standard notation $b = 12 \pi /[11 N - 2 n_f]$ and
denoted $\Lambda = \Lambda_{QCD}$. Usually the argument of $\alpha_s$
in the lhs of Eq.~(\ref{alphahard}) is written without the minus sign
as
$\alpha_s (q^2) = 1/[b \ln (q^2/\Lambda^2)]$ but this
the expression is complemented by the statement that $q^2 < 0$. In
our case $k^2_{\perp}/(1-\beta) >0$, so we keep the minus sign
to show explicitly that the argument of $\alpha_s$ is space-like.
 Obviously,
$\Im \alpha_s =0$ in Eq.~(\ref{alphahard}).
In contrast to it, when the argument of $\alpha_s$ is time-like,
$\Im \alpha_s \neq 0$  and in this case
\begin{equation}\label{alphas}
~\alpha_s (m^2) = \frac{1}{b [\ln (m^2/\Lambda^2) - \imath \pi]} =
\frac{1}{b}\frac{\ln (m^2/\Lambda^2) + \imath \pi}{[\ln (m^2/\Lambda^2) +\pi^2]}.
\end{equation}
When the contributions $ \sim \pi$ in Eq.~(\ref{alphas}) are neglected,
there is no difference between $\alpha_s (m^2)$ and $\alpha_s (-m^2)$.
With the LL
accuracy, $k^2_{\perp}/(1 - \beta) \approx k^2_{\perp}$ and therefore in Eq.~(\ref{ahard})
$\alpha_s \approx \alpha_s (k^2_{\perp})$. The minus sign of the argument of
$\alpha_s$ is traditionally
dropped, which drives us back to the standard expression Eq.~(\ref{aperp}).
Finally, differentiation
over $\ln (Q^2)$ converts Eq.~(\ref{ahard}) into the well-known form:
\begin{equation}\label{msdif}
\frac{\partial M_s}{\partial \ln (Q^2/\mu^2)} = -\frac{1}{2 \pi}
\int_x^1 d \beta\frac{(1- \beta)}{\beta} M(s,Q^2)
\alpha_s \Big(\frac{Q^2}{(1-\beta)} \Big) \approx
- \frac{\alpha_s(Q^2)}{2 \pi}\int_x^1
\frac{ d\beta}{\beta} M(s,Q^2, k^2_{\perp}).
\end{equation}
On the contrary, with the leading logarithmic accuracy,
$\beta_0 \approx k^2_{\perp}/s$ at $x \ll 1$
and
the integration over $k^2_{\perp}$ in Eq.~(\ref{ahard}) runs from $\mu^2$ to $s$.

\section{Parametrization of $\alpha_s$ in the Bethe-Salpeter equations}

In this section we study the parametrization of $\alpha_s$ in the Bethe-
Salpeter equation for the forward scattering amplitude $A$ .
Let us assume that $A$ obeys the following
Bethe-Salpeter equation:

\begin{eqnarray}\label{aeq}
A = A_0 + \frac{\imath}{4 \pi^2}\int d \alpha d \beta d k^2_{\perp}
M ((q+k)^2, Q^2, (s\alpha\beta + k^2_{\perp}))
\frac{sk^2_{\perp}}{(-s\alpha\beta - k^2_{\perp} + \imath \epsilon)^2}
\frac{\alpha_s (s\alpha(1-\beta) - k^2_{\perp})}{[s\alpha(1-\beta) - k^2_{\perp} +
 \imath \epsilon]} \\ \nonumber
 = A_0 + \frac{\imath}{4 \pi^2}\int d k^2_{\perp} d \beta d m^2
 M (s \beta, Q^2,~ (m^2\beta + k^2_{\perp}))
\frac{(1- \beta)k^2_{\perp}}{(m^2\beta + k^2_{\perp} - \imath \epsilon)^2}
\frac{\alpha_s (m^2)}{[m^2 + \imath \epsilon]}.
\end{eqnarray}
The second term in the rhs of Eq.~(\ref{aeq}) is depicted in Fig.~2.
Following Ref.~\cite{ds},
we have replaced the Sudakov variable $\alpha$ by the new variable $m^2 = (p-k)^2$:
\begin{equation}\label{am}
s\alpha = \frac{m^2 + k^2_{\perp}}{1- \beta}.
\end{equation}
$M$ in Eq.~(\ref{aeq}) denotes the upper blob in Fig.~2. It includes both the off-shell amplitude $A$
and a kernel.
We will define $M$ later. Now we just
notice that Eq.~(\ref{aeq}) can be solved only after $M$ has been known. $A_0$
stands for an inhomogeneous term. We do not specify $A_0$ because it
does not affect the $\alpha_s$ -parametrization.
In the first place we focus on integrating  over $\alpha$ in
Eq.~(\ref{aeq}) and introduce

\begin{equation}\label{i}
I = \int_{- \infty}^{\infty} d m^2 M (s \beta, Q^2,~ (m^2\beta + k^2_{\perp}))
\frac{(1- \beta)k^2_{\perp}}{(m^2\beta + k^2_{\perp} - \imath \epsilon)^2}
\frac{\alpha_s (m^2)}{[m^2 + \imath \epsilon]}.
\end{equation}

The integrand of Eq.~(\ref{i}) has the singularities in $m^2$.
First, there are two poles from the propagators:
\begin{equation}\label{mpole}
m^2  = -k^2_{\perp}/\beta + \imath \epsilon
\end{equation}
and
\begin{equation}\label{apole}
m^2 = 0 - \imath \epsilon.
\end{equation}
Second, there are two cuts. The first cut is originated by the
$k^2$ -dependence of $M$. In particular, it can be the logarithmic
dependence. The cut begins at
\begin{equation}\label{mcut}
m^2  = -k^2_{\perp}/\beta + \imath \epsilon
\end{equation}
and goes to the left. The second cut is related to $\alpha_s$. It begins at
\begin{equation}\label{acut}
m^2 = 0 - \imath \epsilon
\end{equation}
and goes to the right. The singularities (\ref{mpole}-\ref{acut})
are depicted in Fig.~3.
The integration over $m^2$ in Eq.~(\ref{i}) runs along the $\Re m^2$ -axis
from $ - \infty$ to $ \infty$, so the integral
can be calculated with choosing an appropriate closed integration contour $C$ and taking residues.
The contour $C$ should include the line $- \infty < m^2 < \infty$ and a
semi-circle $C_R$ with radius $R$. The contour $C_R$ may be situated either in the upper or in the right
semi-plane of the $m^2$ -plane. However, if we choose $C_R$ to be in the upper semi-plane,
we should deal with the cut (\ref{mcut}) of an unknown amplitude $M$, which is
impossible without making assumptions about $M$. Such a closing of the contour was chosen
in Ref.~\cite{ds} but the contribution of the cut (\ref{mcut}) was not
taken into account because there was made the assumption that
\begin{equation}\label{fds}
M(s \beta, Q^2, k^2)  \approx M(s \beta, Q^2, k^2_{\perp}),
\end{equation}
i.e. that $k^2_{\perp} \gg |m^2\beta|$. This assumption made possible
 to calculate the integral $I$ with taking the
residue at the pole Eq.~(\ref{mcut}) where $k^2_{\perp} = |m^2\beta|$.
This obvious contradiction between Eqs.~(\ref{fds})
  and (\ref{mpole})  makes the method of Ref.~~\cite{ds} inconsistent.
Alternatively,
choosing the contour $C_R$ in the lower semi-plane involves analysis of
the cut (\ref{acut}) of  $\alpha_s$ and $\alpha_s$ is known.
By this reason, we choose the latter option for $C_R$. So, as shown in Fig.~4,
the closed contour $C$ includes
the line $- \infty < m^2 < \infty$, the lower semi-circle  $C_R$  and
the contour $C_{cut}$ which runs along both sides of the cut (\ref{acut}).
According to the Cauchy theorem,
\begin{eqnarray}\label{icont}
I_C \equiv \int_C d m^2 K M (s \beta, Q^2, (-m^2\beta + k^2_{\perp}))
\frac{(1-\beta)k^2_{\perp}}{(m^2\beta + k^2_{\perp} - \imath \epsilon)^2}
\frac{\alpha_s (m^2)}{[m^2 +
 \imath \epsilon]} =
 \\ \nonumber
 -2 \pi \imath \frac{(1-\beta)}{k^2_{\perp}}
 M (s \beta, Q^2, -k^2_{\perp}/(1-\beta)) \alpha_s (\mu^2).
\end{eqnarray}
The rhs of Eq.~(\ref{icont}) is
the residue at the pole (\ref{apole}) and  $\mu$ is introduced to regulate
the IR singularity for $\alpha_s$. It should be chosen as large as
\begin{equation}\label{ml}
\mu >> \Lambda
\end{equation}
to guarantee applicability of the perturbative expression in Eq.~(\ref{alphas})
for $\alpha_s$. When the initial
partons in Eq.~(\ref{ima}) are quarks, $\mu$ should also obey $\mu \gg$ the quark mass.
Obviously,
\begin{equation}\label{iccut}
I_C = I + I_{cut} + I_R
\end{equation}
where $I$ is defined in Eq.~(\ref{i}),
$I_R$ stands for the integration over the lower semi-circle and $I_{cut}$  refers
to the integration along the cut (\ref{acut}). $I_R$ can be dropped
because $I_R \to 0$
when $R \to \infty$. Now we specify $I_{cut}$:
\begin{equation}\label{icut}
I_{cut} = -2 \imath \int_{\mu^2}^{\infty} d m^2 M (s \beta, Q^2,~ (m^2\beta + k^2_{\perp}))
\frac{(1 - \beta)k^2_{\perp}}{(m^2\beta + k^2_{\perp} - \imath \epsilon)^2}
\frac{\Im \alpha_s (m^2)}{m^2}.
\end{equation}
The integration in Eq.~(\ref{icut}) cannot be done precisely because it involves the unknown
amplitude $M$ depending on $m^2$. Contrary to the case of the hard kinematics considered in
Sect.~III, the
arguments of the Gribov bremsstrahlung theorem\cite{grib} cannot be used here.
Nevertheless, it is possible to estimate $I_{cut}$. Indeed, the
$m^2$- dependence of $M$ in Eq.~(\ref{icut}) can be neglected in the region
\begin{equation}\label{mreg}
m^2 \ll k^2_{\perp}/\beta.
\end{equation}
Doing so, we obtain the following estimate of $I_{cut}$:
 \begin{equation}\label{icutm}
 I_{cut} \approx -2 \imath \frac{(1 - \beta)}{k^2_{\perp}}M (s \beta, Q^2, k^2_{\perp})
\int_{\mu^2}^{k^2_{\perp}/\beta} d m^2 \frac{\Im \alpha_s (m^2)}{m^2}=
- \frac{2\imath \pi}{b} \frac{(1- \beta)}{k^2_{\perp}}M (s \beta, Q^2, k^2_{\perp})
\int_{\mu^2}^{k^2_{\perp}/\beta} \frac{d m^2}{m^2} \frac{1}{[\ln^2(m^2/\Lambda^2) + \pi^2]}
 \end{equation}
 When $\mu$ is chosen as large that
 \begin{equation}\label{mupi}
 \ln(\mu/\Lambda) \gg \pi/2,
 \end{equation}
we can drop $\pi^2$  in Eq.~(\ref{icutm}) and
arrive at the estimate
\begin{equation}\label{bigmu}
I_{cut} \approx  \frac{2\imath \pi (1- \beta)}{k^2_{\perp}}M (s \beta, Q^2, k^2_{\perp})
\big[ \alpha_s(k^2_{\perp}/\beta) - \alpha_s (\mu^2)\big] .
\end{equation}
Generally, when $\mu$ is not kept very large, though the condition
$\mu \gg \Lambda$ must be satisfied, the integration in Eq.~(\ref{icutm}) issues that
\footnote{A similar estimate was done in Ref.~(\cite{egtalpha}).}
\begin{equation}\label{anymu}
I_{cut} \approx  \frac{2\imath (1- \beta)}{b k^2_{\perp}}M (s \beta, Q^2, k^2_{\perp})
\arctan \Big(\frac{\pi [\ln (k^2_{\perp}/\beta \Lambda^2) -
\ln (\mu^2/\Lambda^2)]}
{\pi^2 + \ln (k^2_{\perp}/\beta \Lambda^2)\ln (\mu^2/\Lambda^2)}\Big).
\end{equation}
Obviously, Eq.~(\ref{anymu}) converts into (\ref{bigmu}) when $\mu$ obeys (\ref{mupi}).
Combining Eqs.~(\ref{icont},\ref{iccut})) and (\ref{anymu}),
we rewrite Eq.~(\ref{i}) as follows:
\begin{equation}\label{ianymu}
I \approx \frac{- 2\imath \pi}{k^2_{\perp}}M (s \beta, Q^2, k^2_{\perp})
(1- \beta) \Big[ \alpha_s(\mu^2) - \frac{1}{\pi b}
\arctan \Big(\frac{\pi [\ln (k^2_{\perp}/\beta \Lambda^2) -
\ln (\mu^2/\Lambda^2)]}
{\pi^2 + \ln (k^2_{\perp}/\beta \Lambda^2)\ln (\mu^2/\Lambda^2)}\Big).
\end{equation}
Assuming  that $\mu$ obeys Eq.~(\ref{mupi}) allows to simplify Eq.~(\ref{ianymu})
down to the very simple expression free of the infrared-dependent contributions:
\begin{equation}\label{ibigmu}
I \approx \frac{- 2\imath \pi}{k^2_{\perp}}M (s \beta, Q^2, k^2_{\perp})
\alpha_s(k^2_{\perp}/\beta).
\end{equation}

\section{Parametrization of $\alpha_s$ in equations for the parton distributions}

Now we can easily consider to the Bethe-Salpeter equations for the
parton distributions.
Indeed,
the parton distributions are proportional to $\Im A$. So, combining Eqs.~(\ref{aeq})
and (\ref{ianymu}) and taking $\Im A$, we arrive at the following expression:
\begin{equation}\label{ima}
\Im A = \frac{1}{2 \pi}
\int^{s}_{\mu^2} \frac{d k^2_{\perp}}{k^2_{\perp}}
\int^{1}_{\beta_0} d \beta  (1- \beta) \Im M ((q+k)^2, Q^2, k^2_{\perp})
\Big[ \alpha_s(\mu^2) - \frac{1}{\pi b}
\arctan \Big(\frac{\pi [\ln (k^2_{\perp}/\beta \Lambda^2) -
\ln (\mu^2/\Lambda^2)]}
{\pi^2 + \ln (k^2_{\perp}/\beta \Lambda^2)\ln (\mu^2/\Lambda^2)}\Big)\Big].
\end{equation}
In Eq.~(\ref{ima}) we have used $\mu^2$ as
the infrared cut-off and the starting
point of the integration over $k^2_{\perp}$.
$\Im M \neq 0$ when $(q+k)^2 >0$.  In terms of the Sudakov variables
the condition $s > (q+k)^2 >0$ can be approximately rewritten as
\begin{equation}\label{beta0}
1 > \beta  > \beta_0 = x +k^2_{\perp}/s.
\end{equation}
 It defines the limits of the integration over $\beta$. It also fixes the upper
 limit of integration over $k^2_{\perp}$ in Eq.~(\ref{ima}). We consider below
 the large-$x$ and small-$x$ situations.

\subsection{Parametrization of $\alpha_s$ in DGLAP}

When $x \lesssim 1$, it follows from Eq.~(\ref{beta0}) that
$\beta_0 \approx x$, so we can write Eq.~(\ref{ima}) in the
DGLAP-like form:
\begin{equation}\label{adglap}
\Im A = \frac{1}{2 \pi}
\int^{Q^2}_{\mu^2} \frac{d k^2_{\perp}}{k^2_{\perp}}
\int^{1}_x d \beta (1- \beta) \Im M ((q+k)^2, Q^2, k^2_{\perp})
\Big[ \alpha_s(\mu^2) - \frac{1}{\pi b}
\arctan \Big(\frac{\pi [\ln (k^2_{\perp}/\beta \Lambda^2) -
\ln (\mu^2/\Lambda^2)]}
{\pi^2 + \ln (k^2_{\perp}/\beta \Lambda^2)\ln (\mu^2/\Lambda^2)}\Big)\Big].
\end{equation}
and express the off-shell $\Im M$ through the off-shell
$\Im A$:
\begin{equation}\label{am}
\Im M ((q+k)^2, Q^2, k^2_{\perp}) = \Im A ((q+k)^2, Q^2, k^2_{\perp})
\frac{1}{\beta}.
\end{equation}

Assuming that $\mu$ obeys Eq.~(\ref{mupi}), allows to approximate the
expression in the square brackets in Eq.~(\ref{adglap}) by the expression
in Eq.~(\ref{ibigmu}). After that we obtain:
\begin{equation}\label{dglap}
\Im A = \frac{1}{2 \pi}
\int^{Q^2}_{\mu^2} \frac{d k^2_{\perp}}{k^2_{\perp}}
\int^{1}_x \frac{d\beta}{\beta} (1-\beta) \Im A ((q+k)^2, Q^2, k^2_{\perp})
\alpha_s(k^2_{\perp}/\beta).
\end{equation}
Approximating $\alpha_s(k^2_{\perp}/\beta) \approx \alpha_s(k^2_{\perp})$
  and adding non-ladder contributions, where the argument of
  $\alpha_s$ is given by Eq.~(\ref{ahard}), leads to multiplying the rhs of
Eq.~(\ref{am}) by the LO DGLAP splitting function(s) which we denote
$P(\beta)$ without specifying.
After that, differentiating  with respect to $\ln(Q^2/\mu^2)$
converts Eq.~(\ref{dglap}) into the well-known DGLAP equation:
\begin{equation}\label{eqdglap}
\frac{\partial \Im A}{\partial \ln (Q^2/\mu^2)} =\frac{\alpha_s(Q^2)}{2 \pi}
\int^{1}_x \frac{d\beta}{\beta}  P(\beta) \Im A (s \beta, Q^2).
\end{equation}
Let us remind that, in general, the parametrization of
$\alpha_s$ for the contributions of the non-ladder
graphs (see Eq.~(\ref{ahard})) considerably differs from the
parametrization of the ladder contributions in Eqs.~(\ref{ima},\ref{dglap}).

\subsection{Parametrization of $\alpha_s$ in small-$x$ evolution equations}

When $x \ll 1$, the lowest limit of integration in Eq.~(\ref{ima}) is
$s\beta_0 \approx k^2_{\perp}$. Also the upper limit
for the $\beta$ and $k^2_{\perp}$ -integrations is $s \beta \approx 1$.
Besides, there can be a kernel $K$ which should not be associated
 with the DGLAP splitting functions. So, in the small-$x$ limit Eq.~(\ref{ima})
 can be written as follows:
\begin{equation}\label{asmallx}
\Im A = \frac{1}{2 \pi}
\int^{s}_{\mu^2} \frac{d k^2_{\perp}}{k^2_{\perp}}
\int^{1}_{k^2_{\perp}/s} \frac{d \beta}{\beta} (1- \beta) K \Im A (s\beta, Q^2, k^2_{\perp})
\Big[ \alpha_s(\mu^2) - \frac{1}{\pi b}
\arctan \Big(\frac{\pi [\ln (k^2_{\perp}/\beta \Lambda^2) -
\ln (\mu^2/\Lambda^2)]}
{\pi^2 + \ln (k^2_{\perp}/\beta \Lambda^2)\ln (\mu^2/\Lambda^2)}\Big)\Big].
\end{equation}

 However, the change of the integration limits does not affect the parametrization
of $\alpha_s$. Indeed, if $\mu$ obeys Eq.~(\ref{mupi}), the expression in the
squared brackets in
Eq.~(\ref{asmallx}) can again be simplified down to  $\alpha_s(k^2_{\perp}/\beta)$,
otherwise it remains as it is in Eq.~(\ref{asmallx}). In contrast to Eq.~(\ref{dglap}),
the small-$x$ Eq.~(\ref{asmallx}) cannot be simplified down to Eq.~(\ref{eqdglap})
with differentiating.

\section{Discussion}
The analysis of the parametrization of $\alpha_s$ we have done in the present paper
can be addressed the wide group of existing and forthcoming evolution equations
of the Bethe-Salpeter type, including BFKL and DGLAP, where one virtual gluon
is factorized out of the blob. Such a gluon can propagate in the $s$-channel as
well as in the crossing channels. We demonstrated that basically the
parametrization of $\alpha_s$ depends on the channel. For the crossing channels,
where the factorized gluons are soft, the parametrization of $\alpha_s$ is
universally given by Eq.~(\ref{ahard}):
$\alpha_s = \alpha_s \big(k^2_{\perp}/(1 - \beta)\big)$. The case of the $s$ -channel, where the
factorized gluon is not soft, is more involved. The effective coupling
$\alpha_s^{eff}$ here
is given by the following expression:
\begin{equation}\label{aeff}
\alpha_s^{eff} = \alpha_s(\mu^2) - \frac{1}{\pi b}
\arctan \Big(\frac{\pi [\ln (k^2_{\perp}/\beta \Lambda^2) -
\ln (\mu^2/\Lambda^2)]}
{\pi^2 + \ln (k^2_{\perp}/\beta \Lambda^2)\ln (\mu^2/\Lambda^2)}\Big).
\end{equation}
However, if the lowest limit $\mu^2$ of
integration over $k^2_{\perp}$ is chosen large enough to obey
\begin{equation}\label{mupiexp}
\mu \gg \Lambda_{QCD} e^{\pi/2} \approx 5 \Lambda_{QCD},
\end{equation}
i.e. when $\mu \gtrsim 50 \Lambda_{QCD}$,
Eq.~(\ref{aeff}) can be simplified:
\begin{equation}\label{aeffpi}
\alpha_s^{eff} \approx \alpha_s(k^2_{\perp}/\beta).
\end{equation}
We obtained Eq.~(\ref{aeff},\ref{aeffpi}) with integrating $\alpha_s (m^2)$
over $m^2$
in the BS equation (\ref{aeq}). Doing so, we accounted for the analytical properties of all
terms in the integrands of the BS equations, which had not been done
in the preceding calculations.
The further simplifications of $\alpha_s^{eff}$ depend on the integration
region over $\beta$. For example, if essentially $\beta \sim 1$,
Eq.~(\ref{aeffpi}) converts into the well-known expression
$\alpha_s^{eff} \approx \alpha_s(k^2_{\perp})$. The effective coupling
(\ref{aeff}) explicitly depends on the value of $\mu$ whereas $\alpha_s^{eff}$
in Eq.~(\ref{aeffpi}) does not contain $\mu$ and looks $\mu$- independent.
However, it also depends on $\mu$, though implicitly, through Eq.~(\ref{mupiexp}). \\
The expression for $\alpha_s^{eff}$ in Eq.~(\ref{aeff}) incorporates the
contributions containing $\pi$.  They are originated by Eq.~(\ref{alphas})
where the analytical properties of $\alpha_s$ are respected. Those contributions
are certainly beyond the logarithmic
accuracy which is typical for the small-$x$ evolution equations,
so there could be a conclusion made that such contributions should be dropped
since very beginning, right in Eq.~(\ref{alphas}).
As the matter of fact, it cannot be done because the BS equations
for the DIS structure functions and parton distributions contain contributions
of the graph in Fig.~2 with the $s$-cut. Obviously, such a cut involves
$\Im \alpha_s (m^2)$,
i.e. $\imath \pi$ -terms in Eq.~(\ref{alphas}). So, neglecting the $\imath \pi$ -terms
before the integration over $m^2$ automatically would mean accounting for the pole contribution
(\ref{apole})  only, i.e.
would lead to fixing $\alpha_s$ at the $\mu$ -scale. However,
Eqs.~(\ref{mupiexp}, \ref{aeffpi})
show that after integrating over $m^2$ the contributions with $\pi$ can be neglected at
large values of $\mu$.

\section{Acknowledgments}
The work is partly supported by the  EU grant MTKD-CT-2004-510126 in partnership
with the CERN Physics Department and Russian State Grant for Scientific School
RSGSS-5788.2006.2.

\newpage
\begin{figure}
\begin{center}
\begin{picture}(144,120)
\put(0,0){ \epsfbox{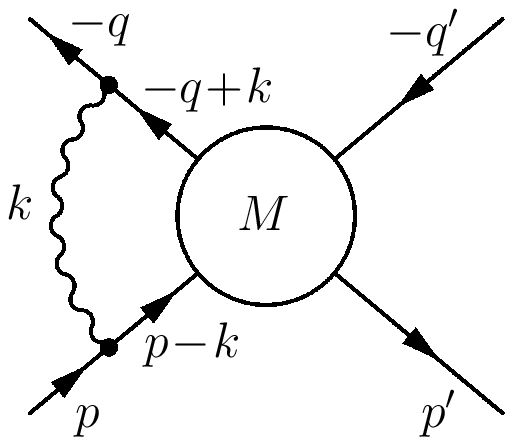} }
\end{picture}
\end{center}
\label{alfafig1}
\caption{The right-hand side of Eq.~(\ref{ms})}
\end{figure}
\begin{figure}
\begin{center}
\begin{picture}(125,150)
\put(0,0){ \epsfbox{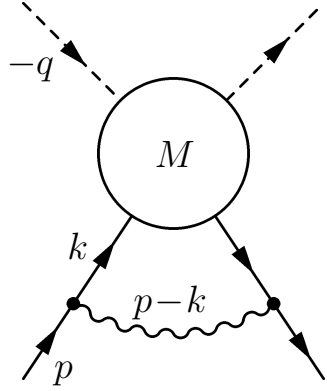} }
\end{picture}
\end{center}
\label{alfafig2}
\caption{The integral contribution in  Eq.~(\ref{aeq})}
\end{figure}
\begin{figure}
\begin{center}
\begin{picture}(180,115)
\put(0,0){ \epsfbox{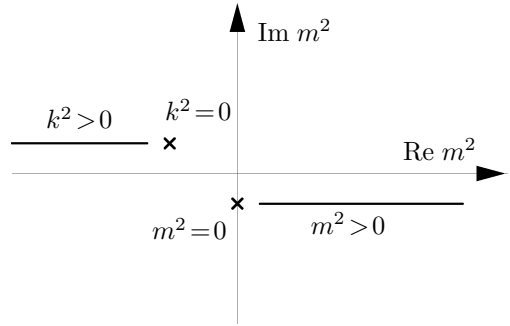} }
\end{picture}
\end{center}
\label{alfafig3}
\caption{Singularities of $I$ given by Eqs.~(\ref{mpole} - \ref{acut})}
\end{figure}
\begin{figure}
\begin{center}
\begin{picture}(180,180)
\put(0,0){ \epsfbox{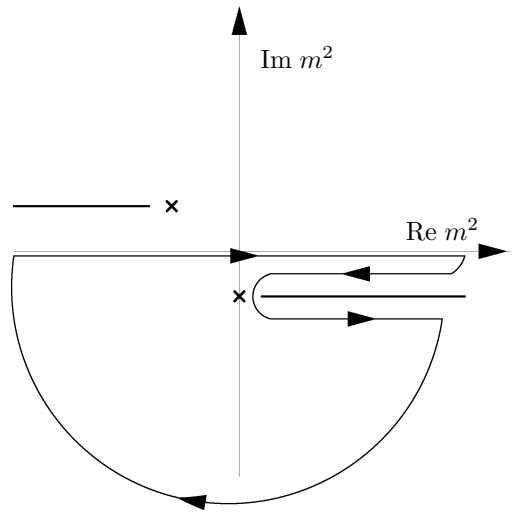} }
\end{picture}
\end{center}
\label{alfafig4}
\caption{The integration contour $C$ for calculating $I_C$}
\end{figure}

\end{document}